\documentclass[runningheads]{llncs}
\usepackage{graphicx}
\usepackage{cleveref}

\usepackage{fancyhdr}
\usepackage{fancybox}
\usepackage{xcolor}

\begin{document}

\title{Global Software Development: Practices for Cultural Differences}
\subtitle{\scriptsize{\textcolor{red}{This is the author-created version of the work accepted at the 19th International Conference on Product-Focused Software Process Improvement 2018 (PROFES 2018). A link to the final authenticated version will be added when it is available.}}}

\author{Marcelo Marinho\inst{1,3} \and
Alexandre Luna\inst{2} \and
Sarah Beecham\inst{3}}

\authorrunning{Marcelo Marinho, Alexandre Luna, Sarah Beecham}

\institute{Federal Rural University of Pernambuco (UFRPE)  Department of Computer Science (DC) Recife, PE, Brazil \\
\email{marcelo.marinho@ufrpe.br}
\and
Federal University of Pernambuco (UFPE) Informatics Center (CIn) Recife, PE, Brazil \\
\email{ajhol@cin.ufpe.br}\\
\and
Lero, the Irish Software Research Centre University of Limerick, Ireland\\
\email{sarah.beecham@lero.ie}}
\maketitle              

\thispagestyle{empty}

\begin{abstract}

Drivers for globalization are significant where today's organizations look for cheaper and faster ways to develop software as well as ways to satisfy quality and investment requirements imposed by customers, shareholders, and governments. Given these needs, Global Software Development (GSD) has become a ``normal'' way of doing business. Working in GSD often require teams of different cultures to work together. A poor understanding of cultural differences can create barriers to trust or missed opportunities. The literature on culture in GSD is either outdated or disparate, requiring practitioners to read many papers to get an overview of how to manage multi-cultural teams. In this study, we aim to highlight how to increase cultural awareness within teams, avoid potential conflict and harness differences for improved team spirit. To answer our research question, ``How should cultural differences be managed, identified and communicated to a GSD team?'', we conducted a systematic literature review of the GSD literature. A synthesis of solutions found in nineteen studies provided 12 distinct practices that organizations can implement, to include, ``provide a cultural knowledge base'', ``understand and make team members aware of cultural differences'' and ``plan responses to mitigate occurrences of cultural misunderstandings''. These implementable cultural practices go some way to providing solutions to managing multi-cultural development teams, and thus to support one of the problem dimensions in GSD and embrace cultural differences.

\end{abstract}

\section{Introduction}

\cfoot{This is the author-created version of the work accepted at the 19th International Conference on Product-Focused Software Process Improvement 2018 (PROFES 2018). A link to the final authenticated version will be added when it is available.}

Software development is a human-centric and socio-technical activity. The complex interaction of different values, attitudes, behavioural patterns, beliefs and communication approaches of members of a project can give rise to misunderstanding and misinterpretation of intent. This misunderstanding can result in conflict, mistrust and under-utilisation of expertise \cite{beecham2015motivates}.

Globalisation implies cultural heterogeneity \cite{noll2010global} requiring organisations to develop high cross-cultural understanding, intercultural communication skills and management competencies. Global organisations that take the cultural context of their teams into account generally experience greater project success \cite{richardson2012process}. Global software development (GSD) is an established field of application and research. Developing software offshore, in different parts of the world, amongst other economic benefits, gives access to skills and knowledge over and above those available in the local environment \cite{beecham2015assessing}.

GSD is widespread, and therefore it is important to have a good understanding of the effects of cultural differences within multi-site teams.  Culture will have a strong bearing on how individuals are motivated and therefore needs to be taken into account when managing, interacting with and rewarding individual software engineers \cite{beecham2015motivates}. The implications of such cultural factors are amplified within global teams increasing project risk and uncertainty \cite{marinho2017managing}.

In managing GSD projects, differences between various dimensions of team members' cultures can lead to increased conflict, reduced quality of cooperation and increased effort required to obtain trust \cite{noll2010global}. Cultural diversity is shown to be an issue in GSD, leading to a significant amount of research on how to manage cultural differences, in fact culture is one of the mature research areas in Global Software Engineering research \cite{ebert2016global}. 

The existing GSD literature also suggests that many software GSD organisations have the potential benefits, resulting in project delays \cite{niazi2016challenges}. For this reason, several studies have looked at methods for improving the results of GSD. Although there has been a significant growth in GSD practice and agile scalability frameworks \cite{razzak2017transition}, this particular strand of research is generally focussed on technical problems and often overlooks cultural difference issues.

Richardson et al. \cite{richardson2012process} developed a global teaming model (GTM) that presented a set of practices for effective management of software teams in GSD scenarios. Although the GTM addresses the need to acknowledge cultural differences and provides some relevant recommendations, the model is focussed on general practices in GSD and does not present clear implementable practices for solving problems arising from cultural differences.  

This paper contributes to the GSD the body of knowledge by presenting a set of practices for managing culture in distributed teams. These cultural practices are the result of a systematic synthesis of recommendations found in the related GSD. We present 12 distilled best practices that address how global practitioners can manage cultural differences in GSD. These practices are aligned with Scaled Agile Framework (SAFe) roles \cite{leffingwell2007scaling}.

This paper is organized as follows: in Section \ref{sec:background}, we introduce the background to the problem and define our research questions. Section \ref{sec:method} describes the method we apply. Sections \ref{sec:results} and \ref{sec:discussion}, present the results, their implications and limitations, respectively. Finally, in Section \ref{sec:conclusions} presents conclusions and future research directions.

\section{Background} \label{sec:background}

\subsection{Global Software Development - GSD}

GSD involves the use of teams located in scattered locations worldwide to develop commercially viable software \cite{beecham2015assessing}. The GSD approach allows companies to leverage development benefits in terms of time, cost and access to appropriate and a wider set of resources \cite{beecham2015assessing}. In developed countries, there is an increasing interest in using GSD to benefit from cost disparities with developing-country labour markets; and some organisations have adopted GSD found a reduction in software development costs and increased product quality as objectives \cite{niazi2016challenges}. 

However, cultural differences associated with geographically distributed teams and the need to work across different time zones are problematic for GSD-based projects \cite{noll2010global}, and several key GSD challenges have been identified, namely, lack of customer involvement, lack of knowledge transfer, hidden costs and communication problems \cite{niazi2016challenges}. Certainly cost advantages are not guaranteed with evidence pointing to increased communication overhead as a costly problem \cite{smite2011whisper}.

The GTM \cite{richardson2012process} has been proposed as a solution to problems in GSD. The GTM was developed as a mechanism for guiding project managers in a global context based on recommendations for addressing time, culture and geographic problems.

\subsection{Global Teaming Model - GTM}

The GTM is a global software engineering model with a particular emphasis on the organisation, governance and management of globally distributed development teams. The specific practices are further elaborated into sub-practices that are used to provide one or more recommendations for implementing detailed actions. Overall, GTM comprises five specific practices, twenty sub-practices and sixty-four recommendations, all of which have been validated against real industrial scenarios \cite{beecham2014motivating,beecham2015motivates}.

Based on their empirical studies, Richardson et al. \cite{richardson2012process} further recommended that ``Cultural differences should be identified and communicated to the management and team members''. However, GTM does not specify how, or by whom, practices should be implemented.

\subsection{Cultural Differences in GSD}

Culture, the sum of the learned values and behaviours shared by a group of people, plays a vital role in guiding how a person performs his/her work through their individual patterns of thinking, feeling and acting \cite{macgregor2005impact}. Given that global virtual team members have diverse national, organisational, professional and cultural backgrounds, cultural diversity is inherent in GSD \cite{richardson2012process}. Studies have shown that cultural diversity can be beneficial in increasing creativity and innovation, which are essential to the knowledge-intensive work of software development \cite{richardson2012process}. Nevertheless, cultural diversity can also become a barrier to communication, coordination and knowledge-sharing and transference, adding difficulties to the management of GSD \cite{noll2010global,monasor2012cultural,niazi2016challenges}.

\section{Method} \label{sec:method}

In carrying out this study, we took a systematic yet focussed approach to examining the relevant literature. Our goal was not to uncover all recorded practices but to select a sufficient collection of studies to enable the identification of recurring themes. 

Established systematic review guidelines  \cite{kitchenham2007guidelines} recommend that a reviewer carry out the following steps: (i) identify the need for a systematic literature review; (ii) formulate review research question(s); (iii) carry out a search for relevant studies. (iv) assess and record the quality of included studies; (v) classify the data needed to answer the research question(s); (vi) extract data from each included study; (vii) summarise and synthesise study results (meta-analysis); (viii) interpret results to determine their applicability; and (ix) write up the study results as a report.

The need for this review was established through an examination of the software engineering literature, which revealed no comprehensive survey addressing the research question regarding practices with respect to managing, identifying and communicating cultural differences in GSD projects.

We therefore sought to answer the following research question: \emph{How should cultural differences be managed, identified and communicated to a GSD team?}

We used the following boolean search string to ensure that we captured a wide variety of papers: \emph{(((Global OR distributed) AND (``software engineering'' OR ``software development'')) AND ``cultural differences'')}.

We used this string to search the metadata relating to journals and conference proceedings in the IEEEXplore, ACM Digital Library, Elsevier ScienceDirect and Scopus bibliographic databases.

\subsection{Document selection}

The search produced 451 references (IEEE = 73; ACM = 58; Scopus = 72; and Science = 248). The idealised selection process had two components: (sp1) an initial selection of research results that could reasonably satisfy the selection criteria (outlined next) based on a reading of the articles' titles and abstracts; followed by (sp2) a final selection against these criteria from the initially selected list of papers based on a reading of their introductions and conclusions.  

\emph{Inclusion/Exclusion criteria}: The following criteria guided the selection of papers that helped us address the research questions. 

We \emph{included}: (i) complete, peer reviewed, published articles; and (ii) empirical studies that fully or partially addressed one or more of our research questions. Texts had to be (iii) published between January 2007 and February 2018. (The start date relates to the year that the Scaled Agile Framework (SAFe) \cite{leffingwell2007scaling} was introduced, and limited the number of studies down to a manageable size).

We \emph{excluded}: (i) texts not published in English; (ii) lacking in proven scientific relevance; (iii) incomplete papers; and (iv) articles that not clearly related to the research questions.

Before accepting a paper into the final set for review, we checked to ensure that there was no replication. For example, if a given study was published in two different journals with a different order of primary authors, only one study would be included in the review; this would usually be the most comprehensive or recent study. Besides, we checked to ensure that there was no duplication. For example, the same paper listed in more than one database, only one study would be included in the review. 

We identified 21 duplicate articles and none replication. After excluding duplicated results from the dataset, we identified 430 articles for inclusion in the initial selection (sp1). Of these, 31 were passed on to the sp2 stage, in which a further twelve were eliminated and nineteen were passed on to the data extraction and data synthesis phase.

\subsection{Study Quality} \label{quality}

The quality assessment criteria adopted for our study are based on principles and good practices established for driving empirical research in software engineering \cite{dyba2007applying}, are briefly summarised as follows. We answered the following questions using \emph{Yes, No, Partially}: (i) Is there a clear definition of the study objectives?; (ii) Is there a clear definition of the justifications of the study?; (iii) Is there a clear definition of how the research was carried out?; (iv) Is validity threat addressed in the study discussion?; and (v) Are the findings of the research clearly stated?

\subsection{Data Extraction}

We examined each selected publication to extract the following elements: (i) study aim or research question; (ii) identified practices for addressing cultural diverse in GSD teams; (iii) other results relevant to the study; and (iv) potential themes emerging from the study's conclusions.

We synthesised the data by first identifying each paper's recommendation as to how to identify and communicate cultural differences to the GSE team. We then generated a summary showing the number of papers mentioning each practice (see Table \ref{practices}). As we gave each occurrence the same weight, the frequencies presented simply reflect how many papers mention a given practice; frequencies therefore reflect prevalence of a theme and not its potential importance.

\section{Results} \label{sec:results}

Of the initial 451 papers examined, 19 met the criteria established in Section \ref{sec:method} as sources for this study; these are listed in Table \ref{tab1}. Analysis of these papers with respect to our research questions revealed practices for identifying and communicating cultural differences in GSD.

\begin{table}
{\scriptsize 
\caption{Paper Accepted for Analysis.}\label{tab1}
\begin{tabular}{|l|l|l|} 
\hline
Author(s) & Title &  Ref\\
\hline
Ayed et al. & Agile Cultural Challenges in Europe and Asia: Insights from Practitioners   &  \cite{ayed2017agile} \\ 

Bannerman et al. & Scrum Practice Mitigation of Global Software Development Coordination  &                         \\ 
                 & Challenges: A Distinctive Advantage?          & \cite{bannerman2012scrum}\\

Breth and  & Toward an Integrative Model of Influence Factors for Success of Global  &                       \\
Drechsler                    &  software development projects                       & \cite{breth2014toward} \\

Boden et al. & Knowledge Management in Distributed Software Development Teams -       &                          \\
             & Does Culture Matter?                                  & \cite{boden2009knowledge} \\

Casey        & Leveraging or Exploiting Cultural Difference?                 & \cite{casey2009leveraging} \\

Chang and B{\'u}rca & An Investigation into how Small Companies in London and the South East   &                             \\ 
                    &  UK Engage in IT Offshore Outsourcing and the Impact of Culture on this  &                              \\ 
                    &  Phenomenon                     & \cite{chang2016investigation} \\

Cramton and Hinds & An Embedded Model of Cultural Adaptation in Global Teams & \cite{cramton2014embedded} \\

Deshpande et al. & Culture in Global Software Development - A Weakness or Strength?       &   \cite{deshpande2010culture}                         \\ 
Dorairaj et al. & Bridging Cultural Differences: A Grounded Theory Perspective & \cite{dorairaj2011bridging} \\

Giuffrida and  & A Conceptual Framework to Study the Role of Communication Through           &                               \\ 
Dittrich                       &  Social Software for Coordination in Globally-Distributed Software Teams    & \cite{giuffrida2015conceptual}     \\ 
Holtkamp et al. & Soft Competency Requirements in Requirements Engineering, Software      &                        \\ 
                & Design, Implementation, and Testing                   & \cite{holtkamp2015soft} \\

Huang and Trauth & Cultural Influences on Temporal Separation and Coordination in Globally     &                         \\ 
                 & Distributed Software Development                  & \cite{huang2008cultural} \\

Mishra and Mishra & Cultural Issues in Distributed Software Development: A Review & \cite{mishra2014cultural}\\

Spohrer et al. & Global Sourcing of Information Systems Development - Explaining Project     &                        \\ 
               &   Outcomes based on Social, cultural, and Asset-Related Characteristics     & \cite{spohrer2012global}    \\ 

Schloegel et al. & Age Stereotypes in Distributed Software Development: The Impact of            &                        \\ 
                 & Culture on Age-Related Performance Expectations & \cite{schloegel2018age} \\

Tugrul et al. & Exploring the Communication Breakdown in Global Virtual Teams & \cite{daim2012exploring} \\

van Marrewijk & Situational Construction of Dutch--Indian Cultural Differences in Global    &                          \\ 
              & IT Projects                                             & \cite{van2010situational} \\

Zaghloul et al & Communication in Firm-Internal Global Software Development with China   & \cite{zaghloul2015communication}  \\ 

Zeid & Using Simulation Games to Teach  Global Software Engineering Courses  &  \cite{zeid2015using} \\ \hline

\end{tabular}}
\end{table}

Each study was assessed independently by two researchers according to five possible quality criteria (see Section \ref{quality}). Only one studies \cite{huang2008cultural} received the maximum score. The other papers were evaluated on the following scales: 12 between 4,5-4,0; 3 between 3,5-3,0 and 3 less or equal 2,5 points.

\begin{table}
{\scriptsize 
\caption{Practices to deal with cultural differences in GSD }\label{practices}
\begin{tabular}{|l|l|l|} 
\hline
Practice & Citations\\
\hline

Understand and be aware of cultural differences &  

\cite{breth2014toward,casey2009leveraging,chang2016investigation,daim2012exploring,dorairaj2011bridging,huang2008cultural,mishra2014cultural,van2010situational,zeid2015using} \\

Make onsite visits &   \cite{boden2009knowledge,chang2016investigation,deshpande2010culture,dorairaj2011bridging,huang2008cultural,schloegel2018age,zeid2015using} \\

Standardise skills required for global team members &   \cite{ayed2017agile,cramton2014embedded,deshpande2010culture,huang2008cultural,schloegel2018age,spohrer2012global,zeid2015using}  \\

Identify and establish the cultural context of each global team &  \cite{casey2009leveraging,cramton2014embedded,daim2012exploring,giuffrida2015conceptual,huang2008cultural,schloegel2018age,spohrer2012global} \\

Provide cultural training &   \cite{bannerman2012scrum,casey2009leveraging,deshpande2010culture,huang2008cultural,schloegel2018age,spohrer2012global} \\

Look out for cultural misunderstanding in Requirement &   \cite{chang2016investigation,breth2014toward,holtkamp2015soft,huang2008cultural,mishra2014cultural,spohrer2012global} \\

Develop and maintain cultural knowledge base &   \cite{breth2014toward,cramton2014embedded,huang2008cultural,schloegel2018age,zeid2015using}  \\

Assign a local manager with the skills needed for a global team &   \cite{boden2009knowledge,chang2016investigation,deshpande2010culture,holtkamp2015soft,huang2008cultural} \\

Offer English language training sessions &  \cite{deshpande2010culture,huang2008cultural,schloegel2018age,zaghloul2015communication} \\

Plan how to mitigate issues caused by cultural misunderstanding &   \cite{bannerman2012scrum,boden2009knowledge,deshpande2010culture,dorairaj2011bridging}  \\

Prepare for distributed meetings &   \cite{dorairaj2011bridging,zaghloul2015communication,zeid2015using} \\

Project managers should take into account cultural differences during &    \\ group exercises &   \cite{chang2016investigation,huang2008cultural} \\ \hline
\end{tabular}}
\end{table}

\subsection{Understand and be aware of cultural differences}

Project managers must recognise and understand the cultural needs of their global teams with respect to their differing organisational, geographic, national, religious, gender and power relations \cite{mishra2014cultural}.

Understanding the challenges associated with cultural differences at an early stage can help workgroups in more effectively managing their cross-cultural communication and conflict management techniques \cite{zeid2015using}.

To address problems in software development while maintaining an atmosphere of mutual respect and team spirit, a global team needs to be aware of its members' respective cultural norms. To develop a better understanding of cultural differences, it is important to consider the personal experiences of team members \cite{dorairaj2011bridging}. In developing an understanding of the subtleties of cultural influence on GSD development, it is also important to draw upon the subjective experience and understanding of the participants \cite{huang2008cultural,van2010situational}.

Because certain activities and types of behaviour that are acceptable in one culture might be unacceptable in another, project managers and participants seeking more effective interaction must understand the multiple individual historical, political, economic, social and cultural contexts of team members \cite{casey2009leveraging}.

The first step to bridging cultural differences is stimulating cultural awareness \cite{dorairaj2011bridging}, and GSD project team members should be prepared for intercultural collaboration by raising an initial awareness of cultural differences \cite{breth2014toward,daim2012exploring,van2010situational,huang2008cultural}.

Cultural awareness helps global teams in collaborating efficiently to achieve the goals and visions of projects \cite{dorairaj2011bridging}. Team members should have an understanding of other members' cultural backgrounds to enable an atmosphere of mutual respect in the software development process \cite{chang2016investigation,dorairaj2011bridging}.

\subsection{Make onsite visits}

Another useful practice is conducting face-to-face meetings at remote sites to provide socialisation opportunities that enhance trust-building \cite{huang2008cultural}.

Some studies \cite{chang2016investigation,deshpande2010culture,dorairaj2011bridging,boden2009knowledge,huang2008cultural} have reported that onsite visits help team members to improve mutual understanding, resolve issues and become familiar with others' work practices, priorities and environments. The results of time spent at outside sites have revealed that personal, face-to-face contact plays a significant role in knowledge exchange \cite{boden2009knowledge,huang2008cultural}.

In addition to building trust in the skills of remote team members, personal meetings play an essential role in learning how to approach offsite personnel. Such face-to-face meetings constitute an essential basis for building social ties, which are reinforced by informal exchange of personal information online \cite{boden2009knowledge,zeid2015using}.

As building an understanding of different values and norms takes some time, management should, despite the additional cost, consider sending team members abroad for more extended periods or even maintaining mixed-cultural teams at project locations \cite{schloegel2018age}.

\subsection{Standardise skills required for global team members}

Spohrer et al. \cite{spohrer2012global} suggested that success on a project level is contingent on the quality of the relationship between both the partner organisations and the respective employee teams. 

Team members and managers encounter cultural differences in terms of expectations, meanings and norms that are stubbornly anchored in local institutions, societies and contexts. By applying an embedded view of cultural adaptation, project managers can become more aware of and effective in responding to the challenges team members face as they bridge worlds \cite{cramton2014embedded,schloegel2018age}. 

Understanding the subtleties of cultural influence on a global project requires drawing on the subjective experiences and understandings of the global virtual team members who are engaged in the work \cite{huang2008cultural}.

Attributes to be considered in understanding other team members include: the language used for communication, which is essential to the interpretation of discussion and communication; gender, which is crucial in identifying the role of female participants as team members while managing cultural diversity \cite{deshpande2010culture}; geography, i.e., where the team members are located; age, as age stereotypes and problems in a GSD context can create considerable barriers to the complex task of software development \cite{schloegel2018age}; Hofstede cultural measures \cite{hofstede1991cultures}, which can help in gaining insight to differing national cultures \cite{deshpande2010culture,zeid2015using,ayed2017agile,schloegel2018age}; and, finally, skills in terms of identifying requirements, developing designs, implementation, testing, leadership and knowledge of other cultures \cite{cramton2014embedded,zeid2015using}.

\subsection{Identify and establish the cultural context of each global team}

Culture has many interacting layers that cannot be examined in isolation. Virtual partners can differ in terms of organisational culture, resulting in work practice differences that impact project success \cite{spohrer2012global,daim2012exploring}.

Different cultures often have different values, attitudes, beliefs, behavioural norms and approaches to communication and problem-solving; these differences can lead to, on one hand, creativity and innovation and, on the other hand, misunderstandings and conflict in the GSD process \cite{schloegel2018age,cramton2014embedded,spohrer2012global,casey2009leveraging}.

Spohrer et al.'s \cite{spohrer2012global} literature and practice investigation revealed that differences in terms of modes of thinking and acting can affect the quality of the relationship between teams. The authors commented that high variance in habits can in some cases be traced back to differences in culture. However, standardised procedures, working practices and roles could decrease misunderstanding and disagreement as to responsibilities. 

Huang and Trauth \cite{huang2008cultural} reported on a global project in which members discussed the effects of organisational culture on values and norms and the effects of bridging some of these cultural differences on time estimation, commitment and adherence to schedule. The project established and promoted global value procedures to leverage its diverse talents and enhance the synergies amongst different company sites. Team members reported that they felt that the organisation valued local cultural diversity while striving to develop a global culture that could serve as a sensemaking device to guide and shape the behaviours of its global workforce. 

Identifying and establishing the cultural context of a global team based on respective social customs can help in understanding organisational categorisations, establishing practices in new distributed teams and re-negotiating practices in established teams \cite{giuffrida2015conceptual}. 

\subsection{Provide a Cultural Training}

Culture and language differences in particular can quickly lead to misunderstandings and/or offense, alienating people and resulting in cooperation barriers \cite{bannerman2012scrum}.

In \cite{spohrer2012global}, a global project with Indian and German teams had an Indian manager at its German site who had been living in Germany for three years. According to the report, this cultural experience provided the manager with an improved understanding of the differences in terms of culture, behaviour and work practices between the two teams.

Training can aid in the education of team members from different cultures or with different religious values and in the interpretation of the behaviours of geographically distinct team members and clients, resulting in an enhanced level of understanding \cite{spohrer2012global,deshpande2010culture,casey2009leveraging,huang2008cultural}.

Schloegel et al. \cite{schloegel2018age} identified cultural training in a GSD context as one of the most important best practices for global teams and posited that it is an economically efficient route to achieving better understanding amongst project participants and, therefore, improved prospects for project success. As a result, cultural diversity training for global team members has become a commonly used strategy for promoting cultural awareness \cite{huang2008cultural}.

\subsection{Look out for cultural misunderstanding in Requirements}

In the early phases of the software development process, many projects must cope with volatile and ambiguous requirements or specifications that can only be resolved through communication amongst all stakeholders. These early phases are, consequently, the most critical of the GSD process \cite{breth2014toward}, and it is necessary that a sufficient understanding of specifications be reached on the vendor side before technical specification of the system begins \cite{mishra2014cultural}.

Chang and B{\'u}rca \cite{chang2016investigation} reported a case study in which there was a misunderstanding of delivery obligations as a result of cultural issues. Following this, the project members worked together to establish social criteria for the delivery of requirements.

In complement, Spohrer et al. \cite{spohrer2012global} and Huang and Trauth \cite{huang2008cultural} reinforce the role of national cultural mindset for the ``problem solving processes'', which in turn, is an essential element of requirements engineering ``to decode the meanings behind each other's language'' about ``business needs'', and transform those needs in software requirements. Hence, Holtkamp et al. \cite{holtkamp2015soft} point out that ``intercultural competences seem to be speciﬁcally important in tasks with a high level of communication''.

\subsection{Develop and maintain a cultural knowledge base}

During the team-building phase, GSD project members require ample time to get to know each other and to reflect on culture-specific communication and collaboration behaviours \cite{breth2014toward}.

Although culturally driven divergences in terms of modes of communication are common, consultation or attempts to obtain more information on other cultures leads to improvements in communication among project members \cite{cramton2014embedded,zeid2015using}

Different cultural backgrounds in GSD projects lead to different team member experiences and knowledge bases; increasing the similarity of the knowledge bases and the extent of a shared understanding will lead to less ambiguity and more successful exchange between project members \cite{breth2014toward}.

Schloegel et al. \cite{schloegel2018age} explained that team members from different national cultures have different communication and problem-solving processes, leading to problems in creating collective knowledge, a shared mental model, social ties and trust, all of which can negatively impact the product quality.

It is therefore vital to create a technical knowledge-base for decoding the meanings behind each member's language, develop a protocol for sharing information and apprising the team of cultural differences and the use of the knowledge-base \cite{huang2008cultural}.

\subsection{Assign a local manager with the skills needed for a global team}

Onsite managers can help manage cultural diversity in GSD \cite{deshpande2010culture}. Local project managers play a critical role in temporal coordination through their responsibility for evaluating available resources, setting realistic goals, monitoring processes, making resource and schedule adjustments when needed and communicating and coordinating with other remote sites concerning changes and progress \cite{huang2008cultural}.

Boden and Liam \cite{boden2009knowledge} found that people with relevant technical and domain knowledge who can connect between cultures make natural facilitators for managing and mediating communication. Similarly, Huang and Trauth underlined the necessity of educating qualified local GSD project managers \cite{huang2008cultural,holtkamp2015soft}.

Chang and B{\'u}rca \cite{chang2016investigation} presented a case in which the local manager became essential to the coordination of cultural challenges. Their finding suggests the importance to the overall GSD process of having a local  manager to mediate communication issues and conduct project activities.

\subsection{Offer English language training sessions}

In situations in which team members do not share a common English language proficiency, training can be provided to enhance language skills \cite{deshpande2010culture}. Such language training can be essential to promoting cultural understanding and facilitating communication \cite{huang2008cultural}.

English language training is best facilitated through the provision of training sessions to all virtual teams with a focus on business terms used in the industry \cite{zaghloul2015communication}. Creating mutual understanding by increasing contact in various ways is even more critical when projects face cultural differences \cite{schloegel2018age}.

\subsection{Plan how to mitigate issues caused by cultural misunderstanding}

Teams with frequent contact and mutual development in which a good professional working relationship is fostered among teams tend to outperform disconnected teams \cite{bannerman2012scrum,deshpande2010culture}.

The project manager should plan responses in the form of shared artefacts and repositories that can mitigate problems caused by cultural differences \cite{boden2009knowledge}. Additionally, the establishment of backup teams at various geographical locations can help international GSD project managers address unforeseen or surplus events that arise as a result of cultural differences \cite{deshpande2010culture}. Global teams should also develop shared work practices to strengthen their team relationships \cite{dorairaj2011bridging}. 

\subsection{Prepare for distributed meetings}

Clear meeting agendas and minutes must be written and disseminated early enough to provide team members with the chance to prepare for meetings \cite{zeid2015using}. Using minutes, team members can write down issues that they wish to discuss during the meeting, enabling them to express themselves clearly and be better understood by other team members \cite{dorairaj2011bridging}.

In general, managers and participants should not provide simple ``yes'' or ``no'' answers. In some cultures it is a common practice for respondents to write up a document outlining their opinions following important sessions. 

It is essential that the project manager repeatedly underline the importance of open conversation and to be understanding of reporting mistakes during projects \cite{zaghloul2015communication}.

\subsection{Project managers should take into account cultural differences during group exercises} 

Cultural differences can occur even when teams share a common language and/or nationality, as differences in ``corporate culture'' can lead to conflicting approaches to problem-solving and communication, which in turn might be misinterpreted as rudeness or incompetence \cite{chang2016investigation}.

Such ``clashes of cultures'' in the form of misunderstandings and schedule delays are common occurrences in global projects. It is therefore necessary to foster an open atmosphere and establish trust relationships at the team level to make team members more willing to raise issues, express concerns and seek and offer help \cite{chang2016investigation,huang2008cultural}. The project manager should repeatedly convey the importance of open conversation. This process can be abetted through the use of a progress-tracking system in which a developer updates the status of his/her task each day to avoid late notifications \cite{huang2008cultural}.

\section{Discussion} \label{sec:discussion}

In this study, we extracted practices for managing, identifying and communicating cultural differences amongst GSD team members. From a directed search of the literature, we have isolated and identified 12 practices that help to align cultural differences amongst members operating in across multi-site and multi-cultural teams.  Research suggests that implementing these practices are important to the success of global software development projects. Such practices can be used in the context of GSD teams that adopt specific approaches by combining well-structured comprehensive methods (traditional) or flexible agile practices or yet the combination between traditional and Agile, a hybrid approach \cite{kuhrmann2017hybrid}; for instance, they can be used in the context of scaling Agile \cite{razzak2017transition}. We believe that this would represent a successful contribution to the Scaled Agile Framework (SAFe).
Although SAFe focusses on large enterprises and takes a scaled approach to Agile adoption, it does not currently represent practices with a cultural difference focus.

To help implement our set of identified practices, we convert each practice into a pattern. We replicate this \emph{patternizing} of a practice method as introduced in Noll et al \cite{noll2014patternizing}. Patterns break down a process into: goals, inputs and outputs, artefacts, and suggested steps. Steps are defined by an action and a role (\emph{who} - a person responsible for carrying out the action (s)). We have defined these roles according to those delineated under SAFe \cite{leffingwell2007scaling}: Development Team (DT); Scrum Master (SM); Release Train Engineer (RTE); Solution Train Engineer (STE); Product Owner (PO); Product Management (PM); and Solution Management (SM).

\subsubsection{Understand and be aware of cultural differences} \emph{Goal}: Understand and raise awareness of cultural differences. \emph{Who}: SM; RTE; STE; \emph{Inputs}: Preliminary understanding of cultural differences; \emph{Outputs}: Preliminary understanding is conveyed to the teams.

\begin{enumerate}
    \item Foster awareness of other cultures to enable problem solving in the software development process for mutual respect and team spirit;
    \item Prepare team members (DT) for intercultural collaboration by making them aware of cultural differences;
    \item Develop an understanding of the subtleties of cultural influences;
    \item Encourage mutual respect of differences to create an environment in which everyone can voice their opinions;
    \item Enable team members to think through the lens of other cultural perspectives.
\end{enumerate}

\subsubsection{Make on-site visits} \emph{Goal}: Provide socialisation opportunities for trust building. \emph{Who}: SM; RTE; STE; \emph{Input}: Initial familiarity with other team members; Project initial understanding and knowledge gained; \emph{Output}: Integration amongst team members.

\begin{enumerate}
    \item Send team members (DT) abroad for more extended periods;
    \item Arrange face-to-face meetings at remote sites.
\end{enumerate}

\subsubsection{Standardise skills required for global team members} \emph{Goal}: Establish guidelines for recruiting and selecting team members who compliment the global team culture. \emph{Who}: PO; PM; SM; \emph{Input}: Organisational guidelines; Project needs; Current profiles of team global members; \emph{Output}: Mapping of team members' skills.

\begin{enumerate}
    \item Identify all global team members in terms of skills, knowledge, experience and behaviours;
    \item Explain to each project team member what is expected from him/her and assess the individual's personal circumstances, motivations, interests and goals.
\end{enumerate}

\subsubsection{Identify and establish the cultural context of each global team} \emph{Goal}: Establish procedures for each global team. \emph{Who}: SM; RTE; STE; \emph{Input}: Cultural information from all involved remote teams; \emph{Output}: Guide for valuing cultural differences and directions for the global project.

\begin{enumerate}
    \item Standardise procedures, working practices and roles;
    \item Identify and establish the cultural context and communication style of each global team based on knowledge of their social customs;
    \item Explain that team members from diﬀerent national and cultural contexts use different communication and problem-solving approaches;
    \item Ensure that cultural differences are valued while developing a global culture that can serve as a sensemaking device to guide and shape the behaviours of the global workforce. 
    \item Establish how each team member is motivated according to the cultural identity.
\end{enumerate}

\subsubsection{Provide cultural training} \emph{Goal}: Ensure cultural diversity in the context of the project is well understood by all stakeholders. \emph{Who}: SM; RTE; STE; \emph{Input}: Cultural information from all involved remote teams; \emph{Output}: Ability to interact with different cultures.

\begin{enumerate}
    \item Create openness based on mutual respect among all individuals and group representatives involved with the project;
    \item Educate team members on cultural and religious value differences to enable mutual interpretation of the behaviour of geographically dispersed team members to attain a required level of understanding.
    \item Explain to individuals that some cultures are more direct than others (e.g. English can be indirect and over-polite when asking someone to do a task, whereas Americans will just tell people what to do and can appear abrupt to some people).
\end{enumerate}

\subsubsection{Look out for cultural misunderstanding in Requirements} \emph{Goal}: Define requirements, according to customer expectations and acceptance criteria. \emph{Who}: PO; PM; SM; \emph{Input}: Cultural information from all remote teams; \emph{Output}: Establishment correct set of requirements and expectations.

\begin{enumerate}
    \item Identify the national cultural mindset of each global team for improvements to communication in requirements;
    \item Clearly and unambiguously define the project results (objectives, deliverables) expected for all parties;
    \item Be aware of the ``mum'' effect dominant in some cultures (hesitation to share bad news).
    \item Be explicit as to which stakeholders' expectations will not be part of the project objectives and the various results or deliverables to be produced.
    \item Be aware of ``false friends'' (words that appear in two languages but have different meanings (for example English magazine and French magasin ``shop'').
\end{enumerate}

\subsubsection{Develop and maintain cultural knowledge base} \emph{Goal}: Establish a system that captures cultural diversity. \emph{Who}: SM; RTE; STE; \emph{Input}: Cultural information from all involved remote teams;\emph{Output}: Structured information on the varying cultures of project teams.

\begin{enumerate}
    \item Collect and register types of behaviour that might be acceptable in one culture but unacceptable in another;
    \item Collect and register technical knowledge to be used at different sites;
    \item Collect information on the interests of the respective parties associated with the project and assess its reliability on a personal and working level;
    \item Incorporate prevailing societal values (as influenced by political opinion, group pressure, interested parties, etc.) that can affect the project;
    \item Develop a protocol for sharing information and making the team aware of cultural differences and the use of the knowledge base.
\end{enumerate}

\subsubsection{Assign a local manager with the skills needed for a global team} \emph{Goal}: Help manage and be responsible for cultural diversity. \emph{Who}: RTE; STE; \emph{Input}: A team member with relevant technical, domain knowledge and people skills; \emph{Output}: Designation of this person as a local leader.

\begin{enumerate}
    \item Designate a team member as a leader (referred to as a coach or scrum master) for bridging cultures.  
\end{enumerate}

\subsubsection {Offer English language training sessions}: \emph{Goal}: To promote cultural understanding and facilitate communication. \emph{Who}: SM; RTE; STE; \emph{Input}: English Language fluency levels of all members in global team; \emph{Output}: Improvement in the linguistic skills of team members.

\begin{enumerate}
    \item Promote English training to enhance team member language skills;
    \item Promote English training with a focus on business terms used in the industry;
    \item Standardise jargon and vocabulary to be used within the project.
\end{enumerate}

\subsubsection{Plan how to mitigate issues caused by cultural misunderstanding} \emph{Goal}: Minimise misunderstanding due to cultural differences and reduce  likelihood of problems arising from cultural differences in the future. \emph{Who}: DT; SM; RTE; STE; PO;  PM; SM; \emph{Input}: Cultural information from all involved remote teams; \emph{Output}: Set of responses for mitigating identified occurrences.

\begin{enumerate}
    \item Get feedback from stakeholders on how they would like to deal with cultural misunderstanding;
    \item Plan responses for mitigating circumstances that occur due to cultural differences;
    \item Establish backup teams at various geographical locations;
    \item Develop work practices and share these with all team members.
\end{enumerate}

\subsubsection{Prepare for distributed meetings} \emph{Goal}: Enable each team member to express themselves clearly and for the other team members to understand what they convey. \emph{Who}: SM; RTE; STE; \emph{Input}: Cultural information from all involved remote teams; \emph{Output}: Improvements in communication and in the relationship among team members.

\begin{enumerate}
    \item Consider the factual arguments around particular issues;
    \item Prepare a presentation that includes rebuttals to possible counter-arguments;
    \item Assess the people who will be involved in the discussion and their likely points of view, interests and relationships;
    \item Prepare an agenda for the meeting in which the issues will be discussed;
    \item Express thanks to the meeting participants for their interest and show appreciation for their input;
    \item Cultivate sustainable relationships with interested parties;
    \item Continuously learn from the experience and apply this learning in the future.
\end{enumerate}

\subsubsection{Project managers should take into account cultural differences during group exercises}: \emph{Goal}: Avoid misunderstandings and scheduling delays arising from cultural differences. \emph{Who}: SM; RTE; STE; PM; \emph{Input}: Cultural information from all involved remote teams; \emph{Output}: Establishment of estimated normalised according to cultural differences.

\begin{enumerate}
    \item Standardise a set of norms for communication and conflict management, for a shared team identity and shared performance expectations;
    \item Create opportunities to stimulate openness amongst the team;
    \item Learn from each situation and continue to improve methods for retaining openness;
    \item Constantly follow verbal and non-verbal cues passed along by the team;
    \item Provide feedback and encourage people to listen.
\end{enumerate}

\subsection{Limitations}

Our search string was intentionally constructed to produce a highly focussed set of candidate papers for review. By including additional terms in the search string (by, for example, adding ``diversity'') and searching additional libraries, we might have produced a larger initial pool of candidates. IEEEXplore, ACM, Science Direct and Scopus comprise a broad array of literature from conferences and journals, ensuring that our findings represent a cross-section of available results. Although our search was focussed, it revealed multiple studies discussing each of the practices for addressing our identified cultural differences in GSD. Although broadening either the set of search terms or target libraries might have revealed additional practices; in our current corpus of papers we started to reach saturation (where on further reading no new themes emerged), and we therefore believe that a larger candidate pool would more likely have only produced additional evidence in support of the practices we have already identified.

\section{Conclusions} \label{sec:conclusions}

It is necessary to take into account many technical, organisational and temporal issues in the interactive and cooperative delivery of GSD solutions, a requirement that is enhanced in situations involving increased team sizes, structures involving teams of teams and more complex management structures.

In such settings, cultural difference can be seen as an enriching factor in which different bodies of knowledge are brought together; it can also lead to severe misunderstanding and conflicts.

As the existing literature did not adequately address particular approaches to successfully implementing practices for managing cultural differences in GSD, we performed an SLR of existing studies to extract specific practices.

Our analysis of these practices revealed a number of actions that organisations can apply in their development processes. In future studies, we will apply the practices identified in this study to a specific organisation with the goal of identifying relevant changes to be taken to enhance the organisation's intercultural effectiveness.

\section*{Acknowledgment}

This work was supported, in part, by Science Foundation Ireland grant no. 13/RC/2094.

\bibliographystyle{splncs04}
\bibliography{mybibliography}
\end{document}